\documentstyle[12pt]{article}
\newcommand{\beq}{\begin{eqnarray}}
\newcommand{\eeq}{\end{eqnarray}}
\begin{document}
\title{Neutrino Masses in a Neutrinosphere}
\author{Leonard S. Kisslinger\\ 
Department of Physics, Carnegie Mellon University, Pittsburgh, PA 15213}
\date{}
\maketitle
\noindent
PACS Indices:14.60.Lm,13.15.+g,14.60.Pq
\vspace{0.25 in}

\begin{abstract}

  The energies of neutrinos in a neutrinosphere, the dense matter created
after the gravitational collapse of a massive star, are estimated.
Cubic equations for energy eigenvalues of neutrinos are used, with the
effective masses found by taking the neutrinos at rest in neutrinosphere matter.
Large differences in the effective mass of some neutrino species in a
neutrinosphere compared to vacuum are found. 
\end{abstract}
\vspace{1mm} 
{\it Keywords~:~}Neutrino masses, neutrino potentials, neutrinosphere

\section{Introduction}

  Neutrino energies and neutrino transport are very important for many aspects
of astroparticle physics and cosmology. Cosmic microwave background (CMB) and
other astrophysical studies have shown that aproximately 23\% of matter in the 
universe is dark matter. Recent measurements of CMB power spectrum favor the
existence of a neutrino species in addition to the three standard 
neutrinos\cite{dh11}, which might be one or more sterile neutrinos associated
with dark matter. An important aspect of neutrino transport is the neutrino
emissivity during the first 20 s after the gravitational collapse of a massive
star, leading to resulting neutron stars with large luminoscity having
large velocities, called the pulsar kick. One possible source or the pulsar
kick is the emission of standard neutrinos after 10 s\cite{hjk07}. Recently
it has been shown that using parameters for electron to sterile
neutrino oscillations, obtained by experiments on active neutrino
oscillations, the pulsar kicks can result from the emission of
sterile neutrinos during the first 10 s, when the neutrinos are in a
neutrinosphere\cite{kj12}.

  Neutrino oscillations, in which one species of active neutrinos converts
to another species, have been studied for tests of symmetry violations.
The effects of matter on neutrino oscillations have been studied for many 
decades \cite{w78,bwpp80}; and matter effects on neutrino oscillations for 
neutrinos traversing the earth have been estimated\cite{os00,ahlo01}. See  
these references for references to earlier work. More recently possible 
tests of time revesal violation\cite{hjk11} and CP volation\cite{khj12}
have been investigated. In all studies of neutrino oscillations the neutrino 
masses are very important, however, only mass differences are needed. 

  More than a decade ago, in preperation for studies of neutrino oscillations, 
energy eigenvalues of neutrinos in the earth were investigated by Kim and 
Sze\cite{ks87} and Freund\cite{freund01} using a cubic  eigenvalue formalism. 
In our present study we use the fomalism of Freund\cite{freund01} to
estmate the energies and effective masses, defined as the energies of neutrinos
with no velocity,  of neutrinos in a neutrinosphere.

\section{Neutrino energy eigenvalues in a neutrinosphere}

  Using the method of Ref \cite{freund01}, the neutrino energy eigenvalues 
$E_i$,
\beq
                 H |\nu_i> &=& E_i |\nu_i> \nonumber \; ,
\eeq
are found as eigenvalues of the matrix $M$\cite{freund01} obtained from the
3 x 3 matrices $U$\cite{freund01} and the Hamiltonian $H$:
\beq
 M=\left( \begin{array}{lcr}s_{13}^2 +\hat{A} + \alpha c_{13}^2s_{12}^2&\alpha
c_{13}s_{12}c_{12}&
s_{13}c_{13} -\alpha c_{13}s_{13}s_{12}^2\\ \alpha c_{13}s_{12}c_{12}& 
\alpha c_{12}^2&-\alpha s_{13}s_{12}c_{12}\\
s_{13}c_{13} - \alpha c_{13}s_{13}s_{12}^2&-\alpha s_{13}s_{12}c_{12}&c_{13}^2
 + \alpha s_{13}^2s_{12}^2 \end{array}
\right)
\eeq

With $c_{ij} \equiv cos(\theta_{ij}), s_{ij} \equiv sin(\theta_{ij})$, and
$\delta m_{rs}^2 \equiv m_r^2-m_s^2$, the parameters in M are:
\beq
\label{parameters}
      c_{12}^2 &=& 0.69 \nonumber \\
      c_{13}^2 &=& 0.9775 \simeq 1.0 \nonumber \\
      \delta m_{12}^2 &=& 7.59 \times 10^{-5} eV^2 \nonumber \\
      \delta m_{13}^2 &=& 2.45 \times 10^{-3} \\
      \alpha &\equiv& \delta m_{12}^2/\delta m_{13}^2 = 0.031 \nonumber \\
      \hat{A} &=& 2E_\nu V/\delta m_{13}^2 \nonumber \; ,
\eeq
with $V$ the potential for neutrino interaction in matter. It is well-known 
that $V= \sqrt{2} G_F n_e$, where $G_F$ is the weak interaction Fermi constant,
 and $n_e$ is the density of electrons in matter. See, e.g., Ref~\cite{os00}.
For neutrinos in earth $V \simeq 1.13 \times 10^{-13} eV$, $\hat{A}\ll 1.0$.

    The eigenvalues of M satisfy the cubic equation (see Eq(17) in 
Ref\cite{freund01} with a=$-I_1, b=I_2, c=-I_3$):
\beq
\label{cubic}  
        \bar{E}_i^3+a \bar{E}_i^2 + b \bar{E}_i +c &=& 0 \\
         a &=& -(1+\hat{A} +\alpha) \nonumber \\
         b &=& \alpha + \hat{A} \alpha c_{12}^2 c_{13}^2 + \hat{A}(c_{13}^2
+\alpha s_{13}^2) \nonumber \\
         c &=& - \hat{A} \alpha c_{12}^2 c_{13}^2 \nonumber \; 
\eeq
with dimensionless quantities $\bar{E}_i = (E_i-E^0_1)/(E^0_3-E^0_1)$, where 
$E_i^0$ are neutrino energy eigenvalues with V=0.
We study neutrinos at rest, so $E_i \equiv m_i c^2 \equiv m_i$, with i = 1, 2, 
and 3; $E^0_i$ are the neutrino masses in vacuum, and $m_i$ are the effective
masses of neutrinos in matter.

From the paramaters  $c_{12}^2$, $c_{13}^2$, and $\alpha$ (Eq.(\ref{parameters}))
one finds
\beq
\label{abc}
     a &=& -(1.031 + \hat{A}) \nonumber \\
     b &=& 0.031 + \hat{A} \nonumber \\
     c &=& -0.0209 \hat{A} \;.
\eeq

We take  $E_\nu \simeq m_3 \simeq \sqrt{\delta m_{13}^2}$, as $m_1 \ll m_3$,
for which $\hat{A}$ is maximum, resulting in the largest matter effect on 
neutrino eigenstates. First, we solve the cubic equations for neutrinos
in vacuum.  From Eqs.(\ref{cubic},\ref{abc}) for V=0 ($\hat{A}=0$)
\beq
\label{Es}
       \bar{E}_1 &=& 3.08 \times 10^{-12} \simeq 0 \nonumber \\
       \bar{E}_2 &=& 0.031 \nonumber \\
       \bar{E}_3 &=& 1.0 \; ,
\eeq 
which is nearly the same as for neutrinos in earth ($\hat{A}\ll 1.0$).

The density of nucleons in the neutrinosphere is approximately that of atomic 
nuclear matter, $\rho_n =4 \times 10^{17} kgm/m^3$. Taking the ratio of the 
electron mass to the proton mass one finds for the electron density in the 
neutrinosphere $\rho_e \simeq 2 \times 10^{11}$ gm/cc, giving the neutrino 
potential in the neutrinosphere  $V \simeq 10^{-2}$ eV. 
.
This gives $\hat{A}_{ns}= 0.404$, which is $\hat{A}$ for neutrinos in a 
neutrinosphere.   Solving Eq(\ref{cubic}) with this dense matter potential 
one finds
\beq
\label{E2s}
       \bar{E}_1^{ns} &=& 0.0208 \nonumber \\
       \bar{E}_2^{ns} &=& 0.3998 \nonumber \\
       \bar{E}_3^{ns} &=& 1.0144 \; ,
\eeq
 
  Comparing Eq(\ref{E2s}) with Eq ({\ref{Es}), $m_3\simeq m_3(V=0)$,
while  $m_2 - m_1(V=0) \simeq 0.4 eV \simeq 13.0 \times (m_2(V=0)-m_1(V=0))$.
Therefore, the neutrino effective masses in the neutrinosphere are
quite different than in earth or vacuum.

\section{Conclusions}

  We find for neurinos in a neutrinosphere during the approximately 10 sec 
after the collapse of a large star the matter potenial is about $10^{11}$ times
larger than in earth, and the effect on the effective masses of neutrinos is 
quite large, with the effective mass difference between $m_2$ and $m_1$ about 
13 times larger in the neutrinosphere than in vacuum or earth. Since the 
neutrino potential for neutrinos in earth is very small, we find that the 
three neutrino effective masses are approximately the same in earth matter as 
in vacuum. The large effect of matter on neutrino oscillations arise from a 
large baseline and depend on the energy of the neutrino beam. 

\newpage

\Large{{\bf Acknowledgements}}\\
\normalsize
\vspace{2mm}

The author thanks Drs. M.B. Johnson and E.M. Henley for helpful discussions.
This work was supported in part by a grant from the Pittsburgh Foundation.

\end{document}